\documentclass{elsart5p}

\usepackage{graphicx}
\usepackage{amssymb}

\usepackage{lineno}
\begin{document}

\begin{frontmatter}

\title{Third Level Trigger for the Fluorescence Telescopes of the
Pierre Auger Observatory}

\author[IPE]{A. Schmidt,}
\author[IPE]{T. Asch,}
\author[IPE]{H. Gemmeke,}
\author[IPE]{M. Kleifges,}
\author[IK]{H.-J. Mathes,}
\author[IPE]{A. Menshikov,}
\author[IK]{F. Sch\"ussler,}
\author[IPE]{D. Tcherniakhovski,}

\address[IPE]{Forschungszentrum Karlsruhe GmbH, Institut f\"ur Prozessdatenverarbeitung und ELektronik (FZK-IPE), 76021 Karlsruhe, Germany}
\address[IK]{Forschungszentrum Karlsruhe GmbH, Institut f\"ur Kernphysik (FZK-IK), 76021 Karlsruhe, Germany}

\begin{abstract}
The trigger system for the Auger fluorescence telescopes is implemented 
in hard- and software for an efficient selection of fluorescence light tracks 
induced by high-energy extensive air showers. The algorithm of the third stage 
uses the multiplicity signal of the hardware for fast rejection of lightning 
events with above 99\,\% efficiency.
In a second step direct muon hits in the camera and random triggers 
are rejected by analyzing the space-time correlation of the pixels. 
The trigger algorithm was tested with measured and simulated showers and 
implemented in the electronics of the fluorescence telescopes. A 
comparison to a prototype trigger without multiplicity shows the 
superiority of this approach, e.g. the false rejection rate 
is a factor 10 lower.
\end{abstract}


\end{frontmatter}

\section{Introduction}

The Pierre Auger Southern Observatory in the Province of Mendoza,
 Argentina in the vicinity of Malarg\"ue measures ultra-high 
energy cosmic rays by an array of 1600 Water-Cherenkov tanks (SD) and 
24 fluorescence telescopes (FD). It combines the surface and the 
fluorescence detection techniques in a hybrid design for minimal systematic errors 
and optimum energy and angular resolution.

Already during construction the Auger collaboration measured the energy spectrum, 
the arrival direction, and the mass composition of cosmic rays above $10^{18}$\,eV 
with high statistical significance. Recently published results \cite{auger_collaboration1} 
shed first light on the astrophysical important questions of possible sources, 
the energy spectrum and GZK-cutoff.

In the following sections the trigger system of the fluorescence telescopes is 
described in general with a particular focus on the software-implemented 
Third Level Trigger (TLT) as part of a multi-level trigger system. We 
specify the selection criteria (cuts) for efficient background rejection and 
measure the figure of merit of this new implementation. 

\subsection{FD system}

The 24 FD telescopes are located in four FD stations around the perimeter of 
the 3000\,km$^{2}$ large SD array. They are designed to measure the fluorescence 
produced by secondary cosmic ray particles due to their interaction with 
$\rm N_2$ molecules in the air. Each FD station comprises 6 telescopes 
with a Schmidt optical system that covers a $30^{\circ}$ azimuth times 
$28.6^{\circ}$ elevation field of view. Fluorescence light falls through a 2.2\,m 
wide aperture with an UV transmitting filter and a corrector lens annulus 
on a spherical mirror and is focused on a photomultiplier camera \cite{FDPaper}. 
The 440 camera pixels consist of hexagonal PMTs (type XP3062) arranged in a 
matrix of 22 rows and 20 columns. 

The required high sensitivity of the PMT to single photons constrains the telescope 
operation to clear nights with dim moonlight. For safety reasons the 
stations are therefore equipped with a sophisticated slow control system 
which allows remote operation of all FD buildings from the central campus (CDAS) 
in Malarg\"ue via a telecommunication network . 

\subsection{FD electronics}
The PMT signals are fed to 20 front-end boards hosted in a
 19'' sub-rack. Each board processes the signals of 22 pixels 
of a single camera column. 
After analogue filtering and gain control at the  First Level Trigger 
(FLT) board the signals are digitized 
with 12-bit resolution and 10 MHz sampling rate. 

The digital data are continuously stored in a $\rm 64K\,\times$ 16-bit memory per pixel, which is
organized in 64 ring-buffers of 1000 words each. If the following trigger stage 
finds a fluorescence light track in the camera image, the data recording continues
after a post-trigger delay of $70\,\mu{\rm s}$ with the next available ring-buffer. 
Otherwise, the current ring-buffer data is overwritten after $100\,\mu{\rm s}$. 
 
\subsection{First Level Trigger (FLT)}

While ADC values are stored in memory, the running sum of the last 10 samples 
is calculated to smooth the random fluctuations of the sky background. 
If the running sum exceeds an adjustable threshold the corresponding pixel
is marked as triggered for a $\rm 20\,\mu s$ coincidence time.

The rate of triggered pixels is measured and kept constant around 100\,Hz by 
regulating the individual thresholds. With this robust regulation scheme we prevent 
increasing random trigger rates despite slowly changing background light intensities 
during the night.

\subsection{Second Level Trigger (SLT)} 

\begin{figure}[ht]
\centering
\includegraphics[width = 0.40\textwidth]{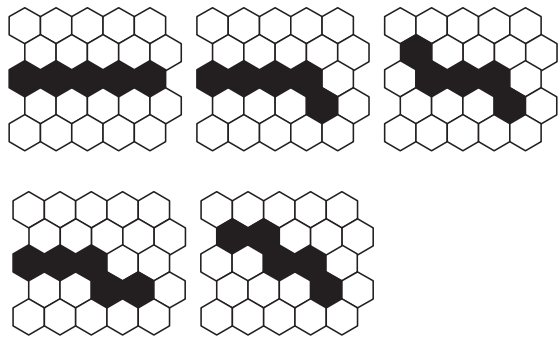}
\caption{Basic topological patterns of triggered pixels as used by the SLT to identify shower track segments}
\label{fig:Pattern}
\end{figure}

While the FLT trigger selects pixels seeing light levels above the common night sky 
background, we have implemented a further stage in hardware identifying tracks of 
triggered pixels in the camera image. For that purpose and for the readout of the data 
the sub-rack is equipped with one Second Level Trigger (SLT) board. It reads the 
triggered pixels from the FLT to construct an image of the camera in its memory.  
Within this image the SLT logic scans for small track segments of 5 adjacent pixels showing
the patterns in fig.~\ref{fig:Pattern} by using highly parallel, pipelined
pattern recognition logic \cite{SLTpaper}. In total, 37\,163
different combinations of pixels are checked every microsecond. If at least 4 
out of 5 pixels forming a pattern have triggered within a $20\,\mu{\rm s}$ long 
coincidence time, the recorded data are kept as event. 

\subsection{Third Level Trigger (TLT)}
\label{sec:old_TLT}
\begin{figure*}[p]
   \hfill
  \begin{minipage}[t]{0.37\textwidth}
      \includegraphics[width = \textwidth]{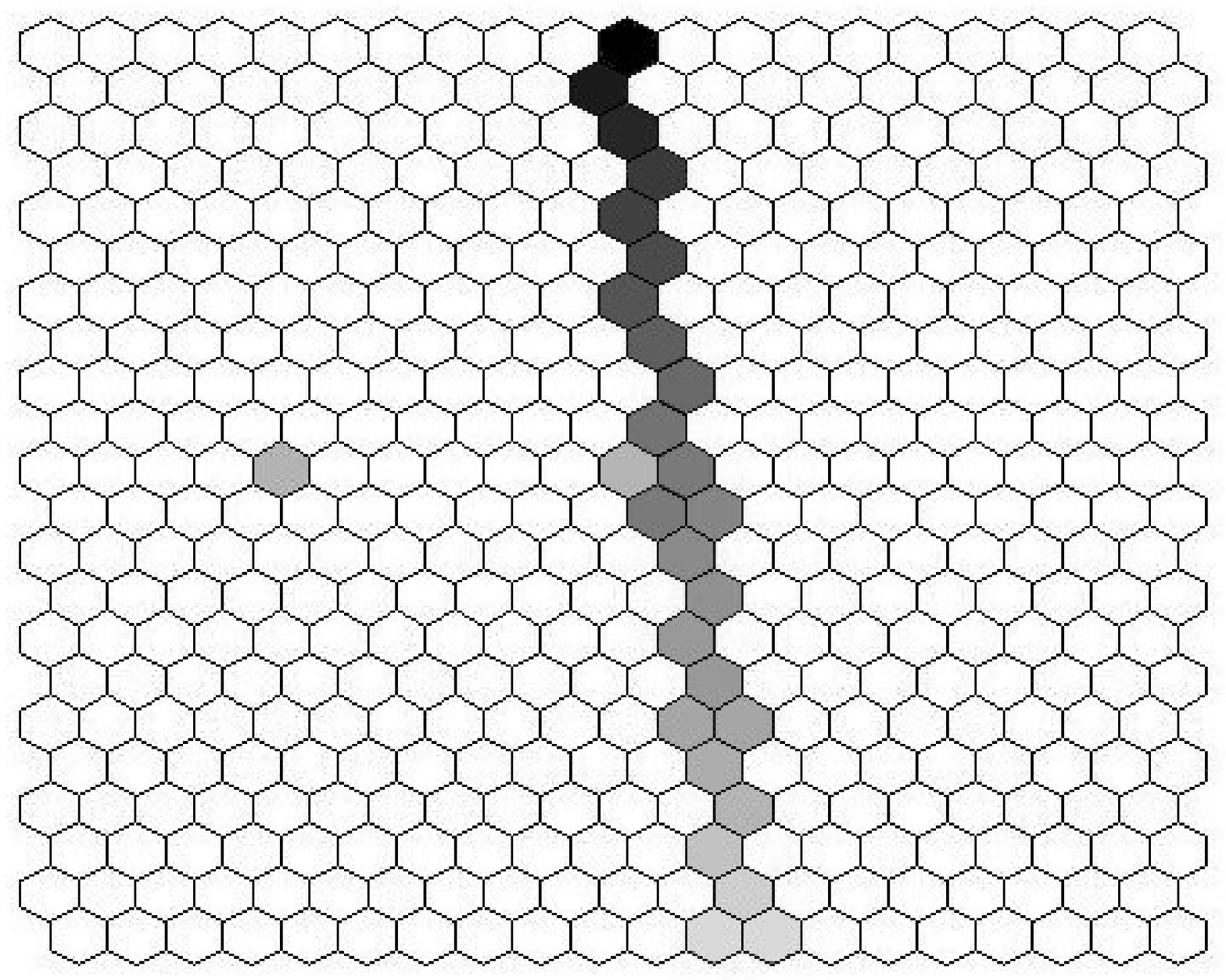}
      \includegraphics[width = \textwidth]{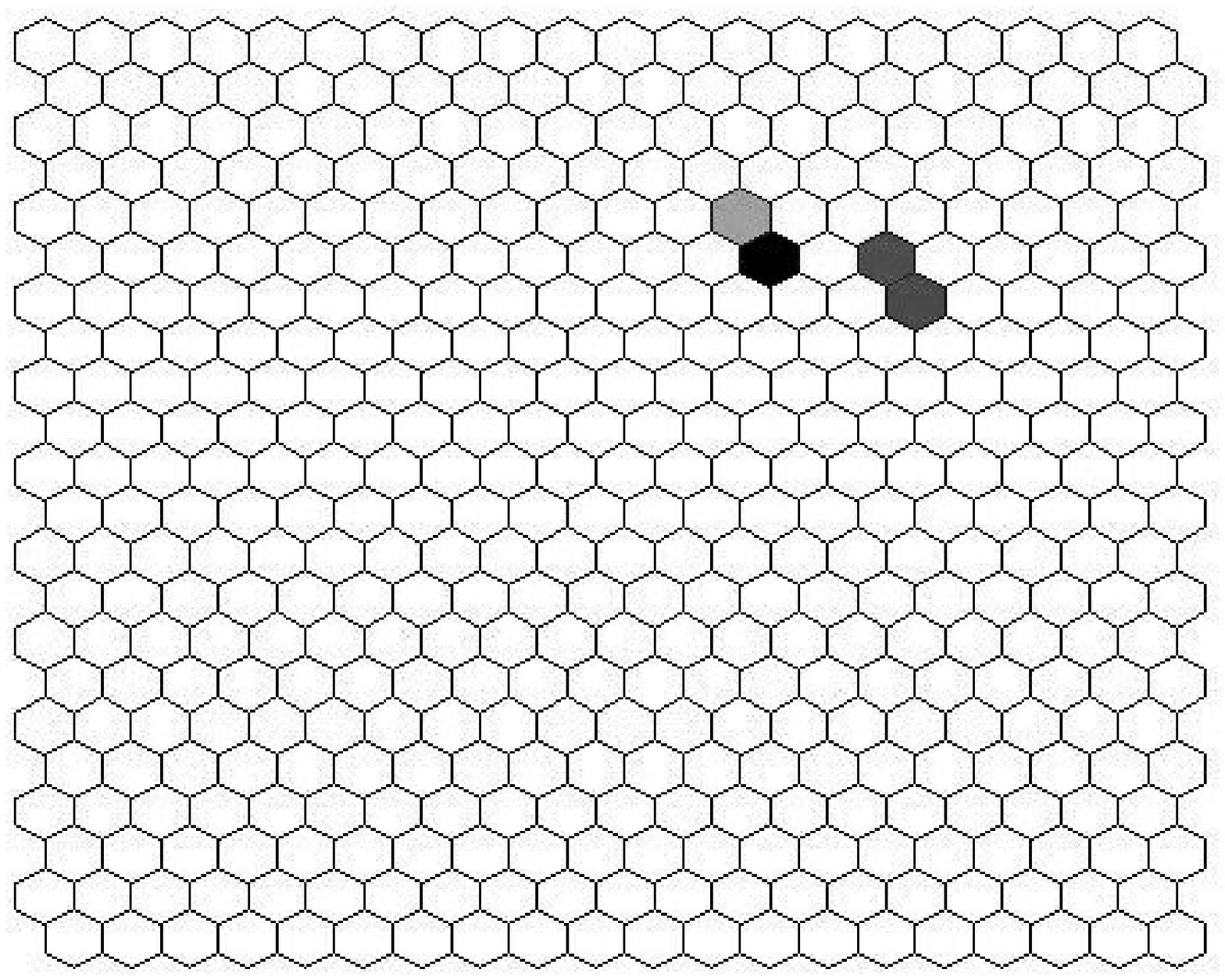}
      \includegraphics[width = \textwidth]{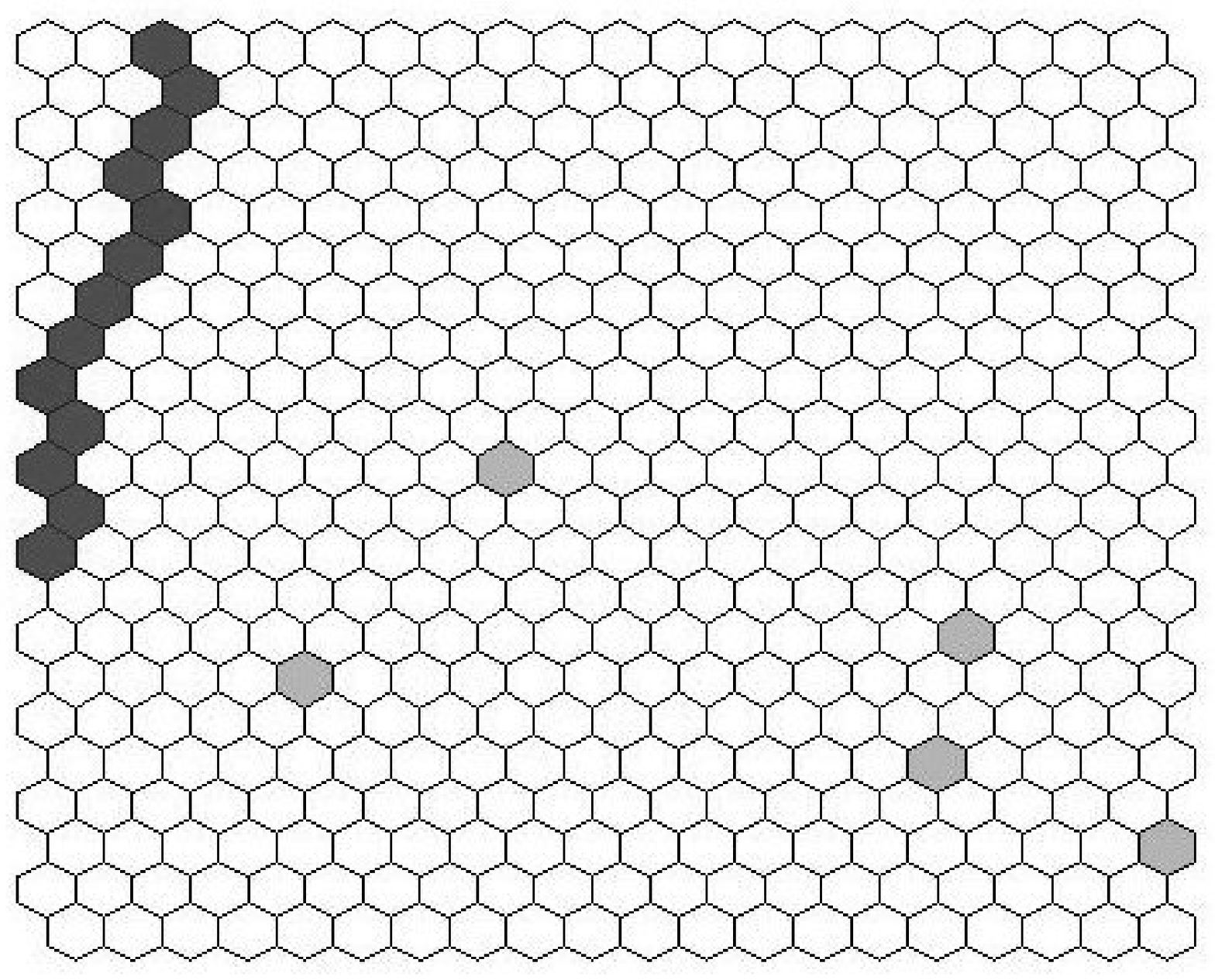}
      \includegraphics[width = \textwidth]{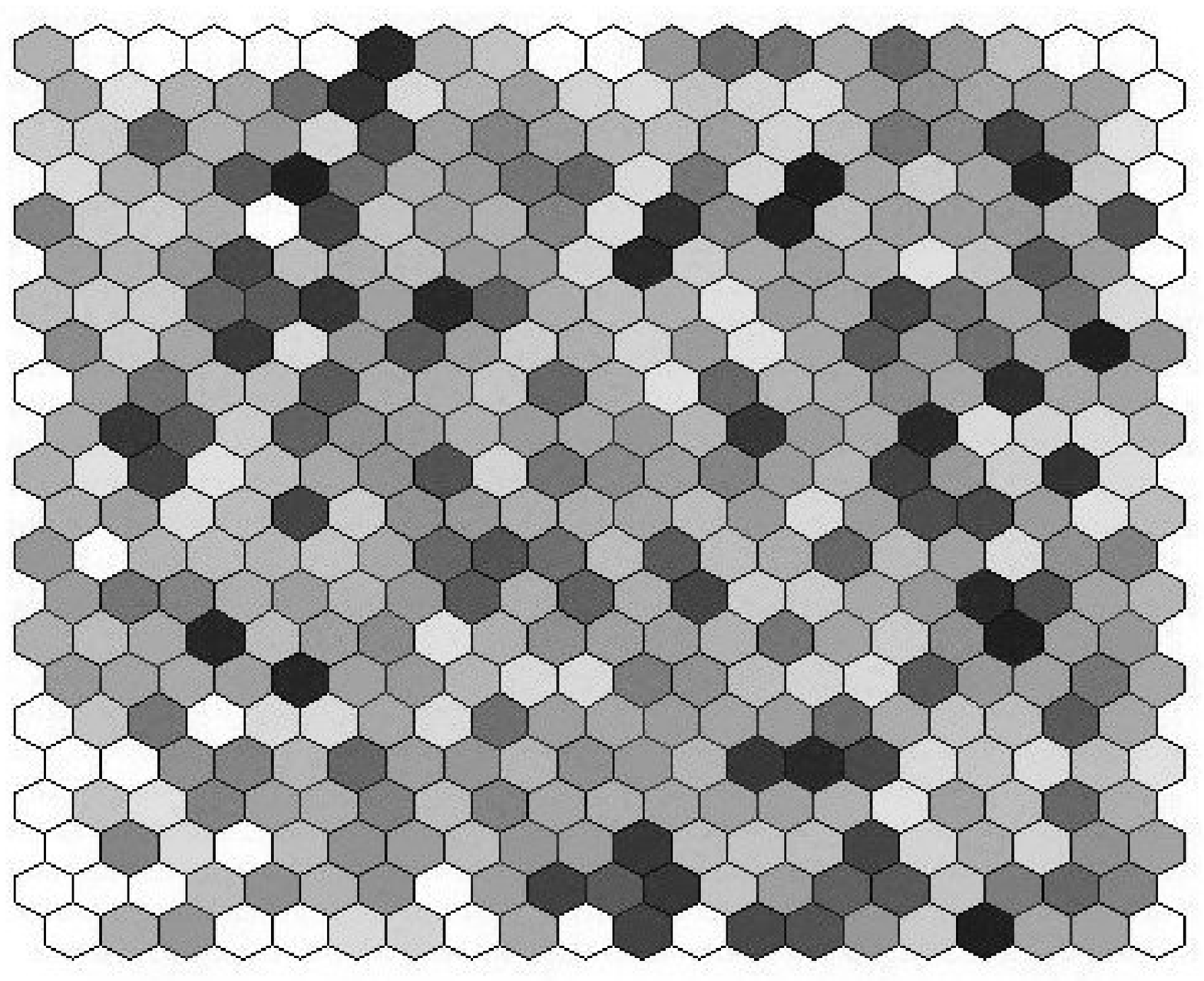}
   \end{minipage}\hfill
   \begin{minipage}[t]{0.40\textwidth}
     \includegraphics[width = \textwidth]{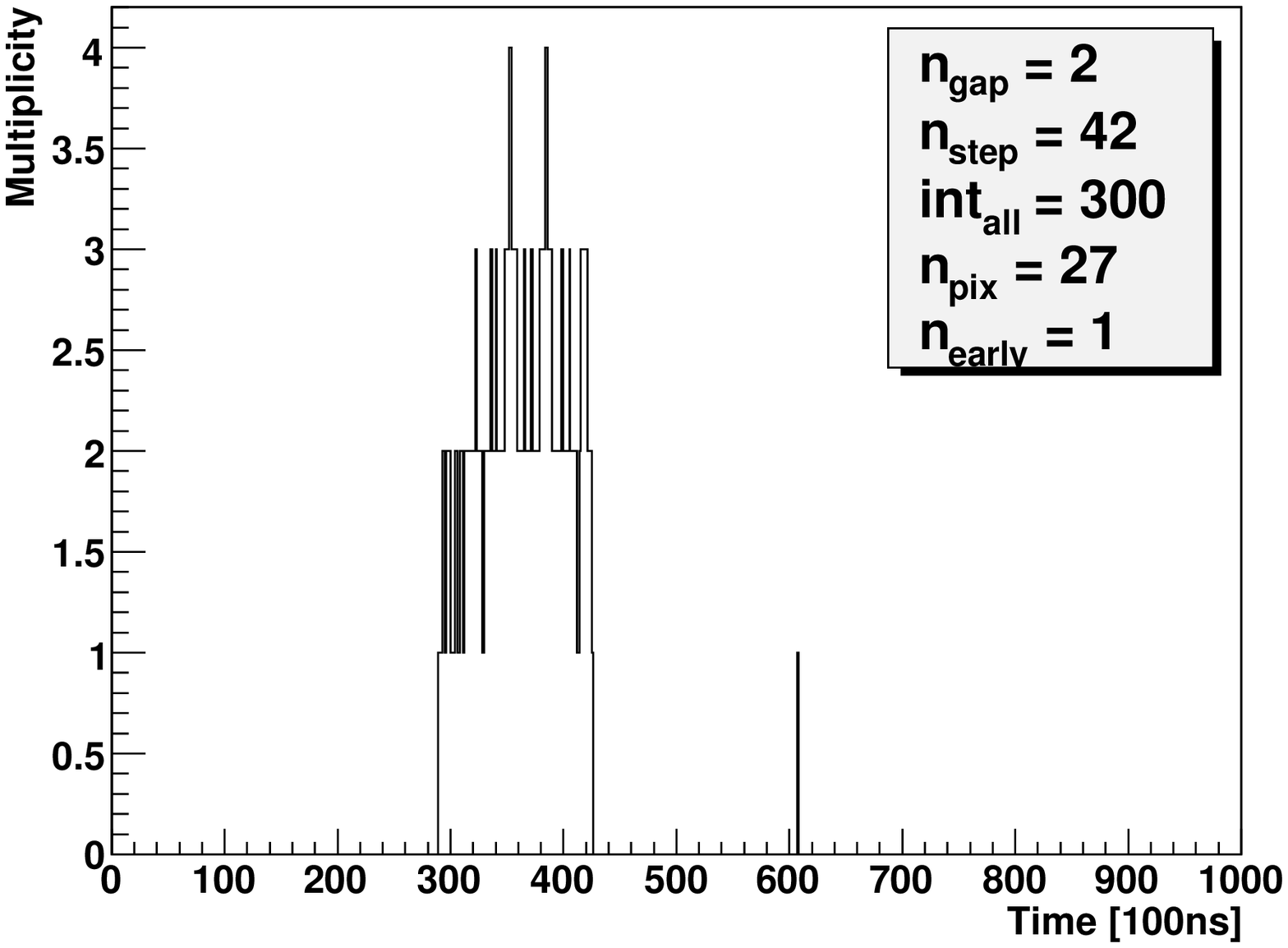}
     \includegraphics[width = \textwidth]{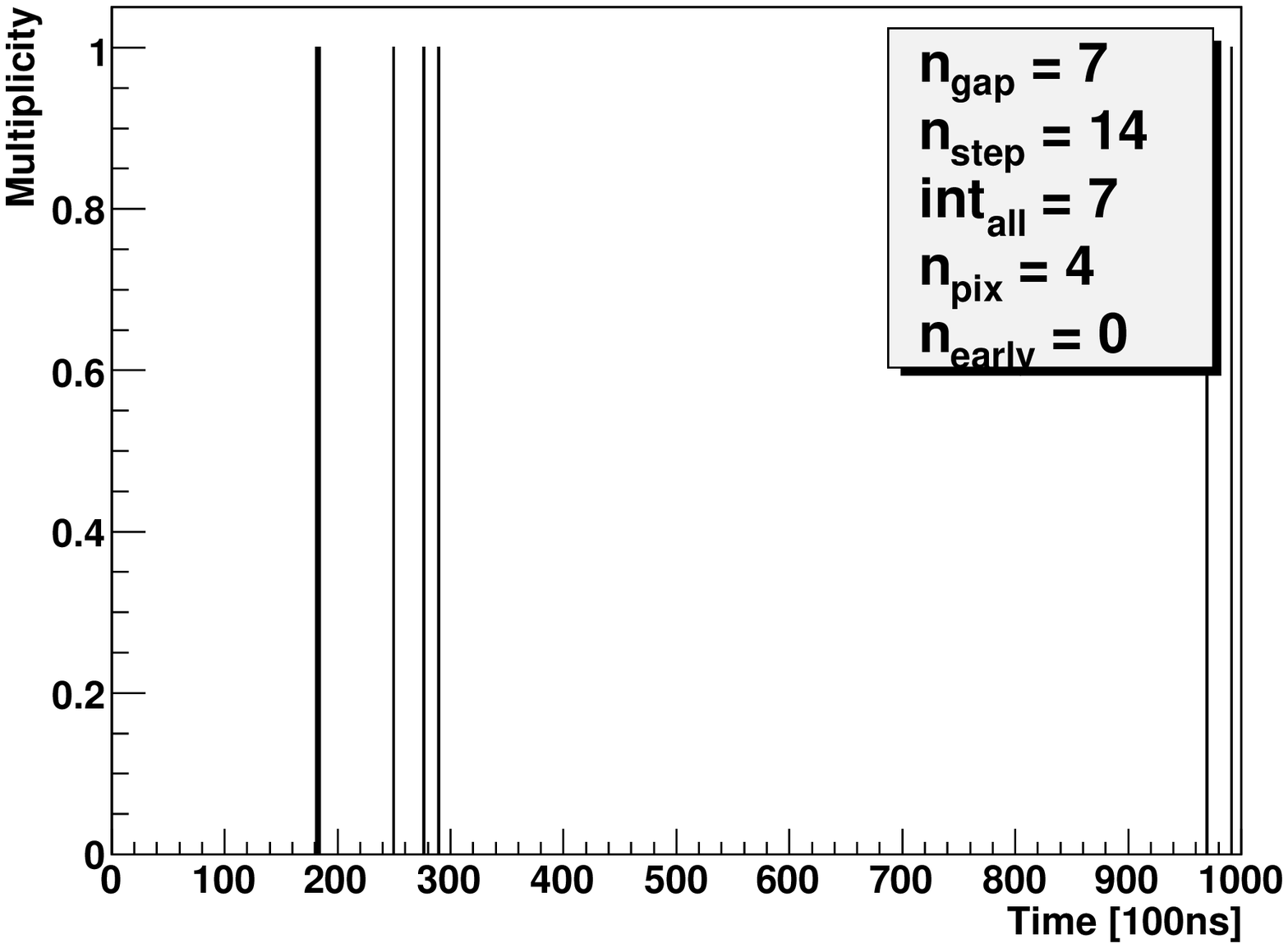}
     \includegraphics[width = \textwidth]{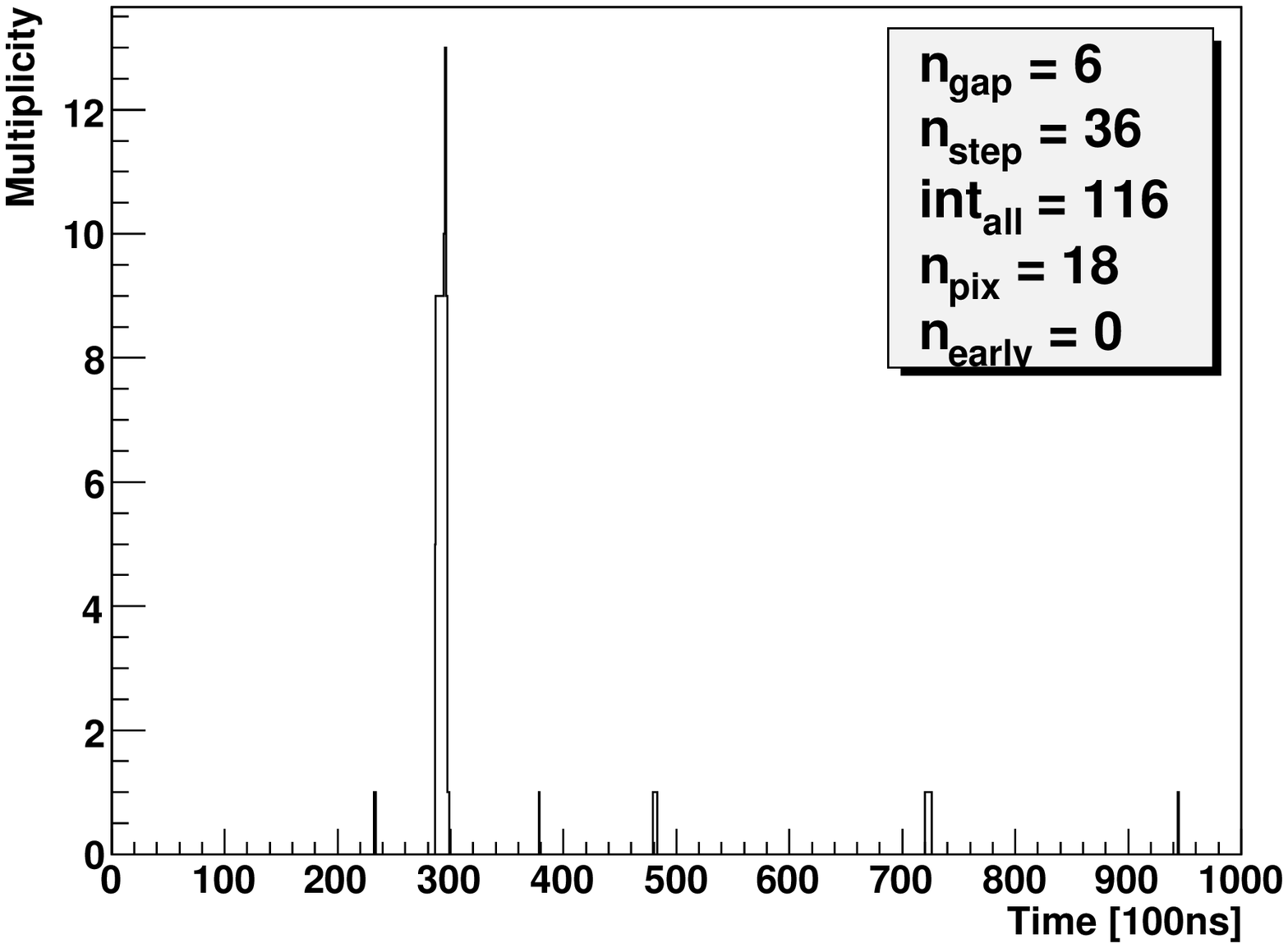}
      \includegraphics[width = \textwidth]{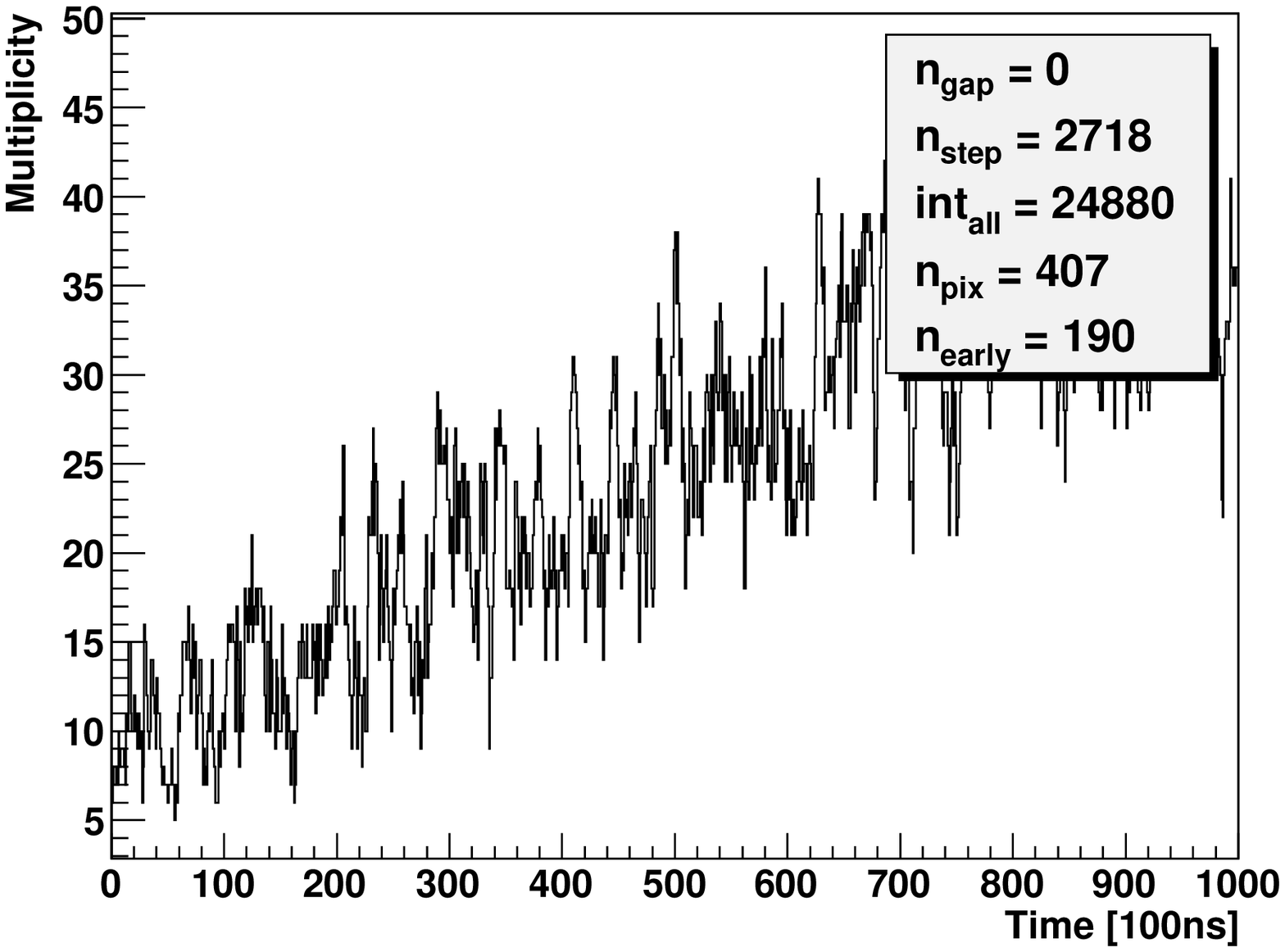}
  \end{minipage}
  \hfill ~
    \caption{Examples of different event types handled  by the third level trigger. 
     \quad \emph{Left panels} show the camera view with 440 hexagonal pixels, signal 
     times are coded by color lightness: a dark pixel corresponds to an early signal.
      \emph{The right panels} give the corresponding multiplicity 
     signal with the multiplicity parameters like explained in section~\ref{sec:multiplicity_parameters}.
    \quad \emph{From top to bottom:} 
    \textbf{a)} Extensive air shower event showing a temporal development in vertical direction, 
    \textbf{b)} random trigger of few pixels, 
    \textbf{c)} camera hit by a cosmic muon resulting in equal pixel times 
    and \textbf{d)} an event caused by strong lightning.}
   \label{fig:example_shower}
\end{figure*}

Different event types pass the first two hardware trigger levels.  
On the one hand there are extensive air showers we want to measure 
(s.~fig.~\ref{fig:example_shower}a), on the other hand different types
of background events.

A small fraction of the background consists of random triggers of few noisy pixels without 
temporal sequence (s.~fig.~\ref{fig:example_shower}b). Another type of background 
are direct hits of muons in the PMTs, that result in tracks of synchronously triggered pixels 
(s.~fig.~\ref{fig:example_shower}c). But the main problem is 
distant lightning causing large areas of the camera to trigger in fast bursts
(s.~fig.~\ref{fig:example_shower}d).

The software implemented third level trigger has to distinguish between background 
and extensive air showers, varying from dim and short tracks for distant showers 
up to an illumination of the whole camera for very close events. 
The key to the trigger decision lies in the timing information of the pixels: Real shower events
always show a temporal development along their track.

The prototype TLT identifies background by analyzing the ADC data
and checking for a temporal sequence inside the 5-pixel SLT-pattern. The major 
disadvantage of this approach was the very time consuming readout of ADC traces,
needed to determine the pixel time. 
Especially during lightning, bursts of events with many pixels were rejected too 
slowly, all FLT buffers were filled and dead time was introduced.

In this paper we will describe a new two-stage TLT. In a first 
step  lightning events are fast and efficiently rejected by cuts on the hardware 
multiplicity data. In the second step the time sequence of the pixels data is
analyzed using their ADC data.

In the DAQ chain the TLT is followed by the Event Builder, which merges concurrent
events in neighboring telescopes together. A subsequent algorithm tries to 
calculate the preliminary shower direction to form a hybrid trigger (T3) for the suface detector.
Finally the event is stored on hard disk and transmitted later on to the central campus via
microwave uplink.

\subsection{Data sample}
\label{sec:data_sample}
For the developement of the new trigger we need a set 
of background events and extensive air showers. Therefore we randomly 
selected 8839 events from one year of data taking and manually separated 
them into 2425 shower and 6414 background events. This 
classification was done very conservative to prevent the misinterpretation of potential 
showers as background. The separation was successfully cross-checked by a reconstruction 
with the detector simulation and analysis framework Offline\cite{offlinepaper}. 
None of our events classified as background, but 25\,\% of our shower
candidates can be reconstructed. This confirms the conservative character of our manual
classification.

After using this data sample for the trigger developement, we determine the background
rejection efficiency with a second batch of background events independentely from the
training data. The acceptance of shower events, especially
at high primary energies, is checked with Monte-Carlo-simulated showers.

\section{Multiplicity based trigger}
In order to speed up the trigger decision, the new TLT has to abstain from
reading out ADC traces until the data load is sufficiently reduced. 
% Anyhow the temporal sequence of the triggered pixels has to be analyzed.
As the digital electronics of FLT and SLT is based on flexible re-programmable
FPGA logic it was easy to update the firmware to provide a hardware 
calculated multiplicity. 

The multiplicity is the number of simultaneously (i.e. within 100\,ns) triggered camera pixels.
A pixel contributes to the calculation of the multiplicity as long as its 
running ADC-sum exeeds the FLT threshold. Contributing pixels of one column are 
counted at the FLT, the total sum for the full camera is evaluated at the SLT. 
The hardware implementation limits the multiplicity to a maximum value of 63.

The chronological sequence of the multiplicity values with full 100\,ns time resolution
describes the temporal development of the overall camera picture. 
This \emph{multiplicity signal} is directly accessible from the hardware and therefore 
particular suitable for effective and fast background discrimination.

\subsection{Multiplicity parameters}
\label{sec:multiplicity_parameters}
In order to capture the characteristics of the multiplicity signal, 
the following 5 parameters were selected, distinguishing well between 
background and shower events:

\begin{list}{}{\setlength{\labelwidth}{4em}\setlength{\leftmargin}{3em}}
\item[$\bf n_{gap}$] Number of gaps in the multiplicity signal, 
i.e. the number of periods without triggered pixels.
\item[$\bf n_{step}$] Sum of steps in the multiplicity signal; 
the absolute values of steps both up and down are added.
\item[$\bf int_{all}$] Integral of the multiplicity signal over the full event window of 1000 values. 
It is equal to the number of triggered pixels times their trigger duration.
\item[$\bf n_{pix}$] Number of triggered pixels. 
This is not a multiplicity parameter, but an overall characterization of the event.
\item[$\bf n_{early}$] Number of pixels, which trigger in the first 5 $\mu$s 
of an event. This value is no real multiplicity parameter either.
\end{list}
We developed 5 independent cuts on these parameters using the  
sample of manually classified events mentioned in sec.~\ref{sec:data_sample} 
as training data set.

\subsubsection{Cut 1}
The SLT divides the readout window in a $30\,\mu$s pre-trigger and a $70\,\mu$s post 
trigger window. Only few pixels should trigger in the first
microseconds of each event. However during lightning the general variance of the 
ADC-traces is such high, that the random pixel trigger rate increases strongly.
Hence we keep only events with less than 6 pixels triggering within the first 
$5\,\mu$s of the event:
\begin{equation}
\rm n_{early} < 6
\label{eq:cut1}
\end{equation}
This constraint already rejects 51.7\,\% of the background in our data sample 
without rejecting any showers.

\subsubsection{Cut 2}
Far away showers cause long pulses in the ADC traces, as it takes them longer to 
pass the field of view of a single pixel. E.g. a vertical shower at 10\,km distance 
illuminates each pixel for about $0.9\,\mu$s. In contrast close showers have shorter, 
but much higher pulses. As the FLT uses a boxcar sum over 10 values for the trigger, 
a short and high peak leads also to a trigger lasting 10 clock cycles.

Therefore showers provide independently of their distance a minimum pixel trigger
duration of about 1 $\mu$s while background events frequently consist of many short 
pixel triggers. 

The integral of the multiplicity signal $\rm int_{all}$ is the sum of all
pixels times their trigger duration. The cut 
\begin{equation}
\rm int_{all} > 8 \cdot n_{pix} -100
\label{eq:cut2}
\end{equation}
rejects all events where a lot of pixels trigger for less than $0.8\,\mu$s. 
28.3\,\% of the background is rejected by this cut without losing any shower event

\subsubsection{Cut 3}
In shower events pixels trigger one after the other, while lightning causes rather 
independent pixel triggers. For this reason the multiplicity signal of showers 
usually consists of continuous areas, while the signal of background events is 
interrupted by gaps. 

By accepting only events with
\begin{equation}
\rm n_{gap}<40
\label{eq:cut3}
\end{equation}
we reject 27.7\,\%  of the background.

\subsubsection{Cut 4}
Sometimes strong lightning events have so many pixel triggers, that they form a continuous
area in the multiplicity signal by chance. These events will not be rejected by 
cut~(\ref{eq:cut3}). However, as the fluctuation of the multiplicity signal is higher than for 
shower events, a cut on $\rm n_{step}$ can identify such events. 
\begin{equation}
\rm n_{step} < 2\cdot (n_{pix} + 25)
\label{eq:cut4}
\end{equation}
This constraint rejects all events, where more than 25 pixels trigger twice, 
which affects 33.1\,\% of the background events.
Only one noisy shower candidate is rejected.

\subsubsection{Cut 5}
During very intense lightning the multiplicity signal gets saturated at its maximum 
value of 63. In the saturated region the multiplicity is constant and the evaluated 
value of $\rm n_{step}$ is getting  lower. Thus cut~(\ref{eq:cut4}) does not work. Such events
are identified by a cut on large values of $\rm int_{all}$:
\begin{equation}
\rm int_{all} < 10000
\label{eq:cut5}
\end{equation}
9.5\,\% of the background is rejected, no additional shower event is lost.

\subsubsection{Results of the multiplicity trigger}
The 5 cuts widely overlap each other reducing their combined rejection efficiency.
In total 58.2\,\% of the background in our data sample is rejected.

The cuts reject background independently from each other, so their order does not
matter in terms of rejection efficiency. So the fastest and most efficient cuts 
are performed first in order to reduce data load for the next steps.

\begin{figure}[ht]
\centering
\includegraphics[width = 0.43\textwidth]{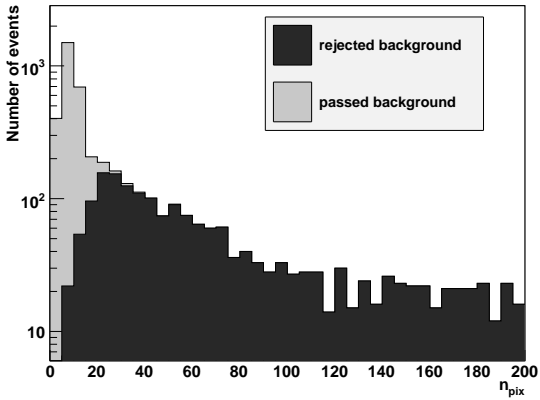}
\caption{Distribution of the background event size in our data sample and rejection
of this background.}
\label{fig:rejection_mcuts}
\end{figure}

The fraction of background event rejected by the multiplicity cuts strongly 
depends on the number of triggered pixels $\rm n_{pix}$: While 99\,\% of the 
large background events with $\rm n_{pix}\ge 25$ are rejected, only 7\,\% of the
smaller events are caught (s.~fig.~\ref{fig:rejection_mcuts}). 

Although cut~(\ref{eq:cut2})\,-\,(\ref{eq:cut5}) contribute only by a
few percentage points to the overall rejection efficiency, they play an important 
role to increase rejection of large background events up to 99\,\%.

Unfortunately it is not possible to enhance the efficiency at low pixel numbers with
additional multiplicity cuts, the multiplicity information is not detailed enough.
For example the signature of a muon hit (short and high ADC-Pulses at the same time)
fades out in the multiplicity signal, as the trigger time is extended 
to 1 $\mu$s by the boxcar sum of the FLT, and only the sum of all pixels is displayed.

However, the critically high data load from bursts of events with high pixel numbers 
during lightning is eliminated in a fast and efficient way.

\section{Analysis of the space-time correlation}
For the reduced data stream of events with few pixels it is now feasible to
readout complete ADC traces in order to improve rejection of muon hits and 
random triggers without deadtime. For this purpose we developed an algorithm that analyses the
temporal sequence of the triggered pixels with full 100\,ns time resolution in 
4 steps:

\subsubsection{Finding the shower center}
At first we have to remove disturbing pixels aside from the real shower track.
Scanning the matrix of triggered pixels row by row, the algorithm finds groups 
of connected pixels (allowing maximal 1 gap). The group with the largest number 
of pixels is regarded as forming the "shower center" which is used in the
subsequent analysis.

\subsubsection{Direction of the shower center}
\begin{figure}[ht]
\centering
\includegraphics[width = 0.43\textwidth]{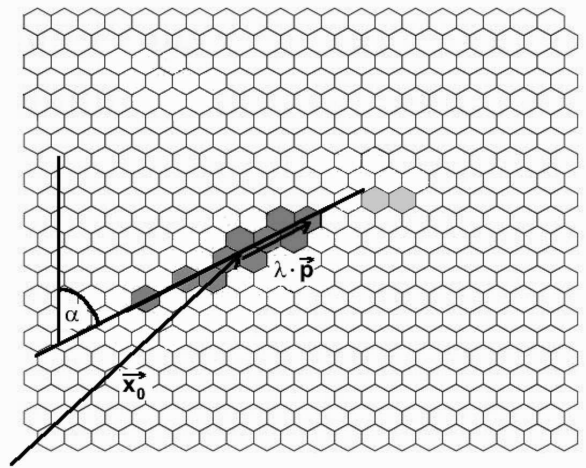}

\caption{Example of a shower event with straight line fit. Both light grey
pixels are not taken into account as they do not belong to the shower 
center.}
\label{fig:showercore1}
\end{figure}

A straight line parameterized in the form (\ref{eq:straightline}) is 
fitted to the pixels of the shower center using the least square method 
(s. fig.~\ref{fig:showercore1}). 
For each pixel its line position $\lambda$ is calculated and the pixels are sorted by
rising $\lambda$-values.
\begin{equation}
\vec{x}=\vec{x}_0+\lambda \cdot \vec{p} \quad \quad 
{\rm with} \quad\quad \vec{p}\,={\sin \alpha \choose \cos \alpha}
\label{eq:straightline}
\end{equation}

\subsubsection{Time assignment}
The pixel times must be determined by analyzing the ADC data. The 
FLT-pixel time is not suitable, as the time of threshold crossing 
is blurred by the 10 entry running sum.
Good results are achieved by a simple maximum search of a running
2-bin ADC-average. As we can easily access the second level trigger time with a 
low resolution of $1\,\mu$s, the time assignment can be accelerated by constraining 
the maximum scan to a $2\,\mu$s window around the SLT time. 

The number of entries in the moving average is  a compromise between best noise suppression and
acceptable time-blurring. Another compromise is the width of the search window: a wider window around the 
SLT time might find a higher ADC pulse, but increases the danger of finding wrong pulses. 
Both values were varied and tested, above listed values are used.

Randomly triggered pixels are removed from the shower center group, if their pixel time
differs by more than 2 rms deviation from the average shower center time.

\subsubsection{Check for temporal sequence}
Finally the algorithm runs through the list of $\lambda$-sorted pixels to count 
how often the time difference is positive (increment of $\rm n_{up}$) or negative 
(increment of $\rm n_{down}$). If the time of adjacent pixels is equal, that pixel is 
not taken into account. The defined temporal propagation of showers requires either always 
increasing or always decreasing pixel times. Thus the absolute value $\rm |n_{up} - n_{down}|$ 
is a good criteria for shower events independent from their direction.

A single randomly triggered pixel has big influence on the temporal sequence of showers
with few pixels. Therefore shower centers with less than 7 pixels are checked again to verify,
if their value of $\rm |n_{up} - n_{down}|$ increases when one pixel is left out of 
the $\rm n_{up}$, $\rm n_{down}$ calculation.

\begin{figure}[ht]
\centering
\includegraphics[width = 0.43\textwidth]{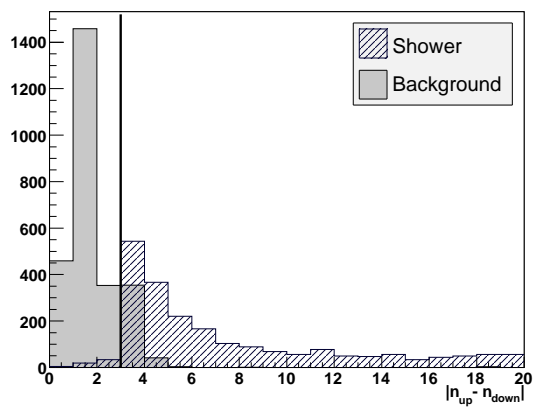}
\caption{Histogram of the $\rm |n_{up} - n_{down}|$ values of all events
passing the multiplicity cuts. Events to the left of the vertical line
are rejected}
\label{fig:stc-cut}
\end{figure}

Figure~\ref{fig:stc-cut} displays the distribution of the parameter $\rm |n_{up} - n_{down}|$
for shower and background events. The cut
\begin{equation}
\rm |n_{up}- n_{down}| \ge 3
\label{eq:stc-sut}
\end{equation}
rejects events with only a weak temporal sequence.

This last cut increases the background rejection up to 93.6\,\%, but it also rejects 2.4\,\%
of our shower candidates. Those 56 falsely rejected shower events were checked in detail:
The majority of them could not be reconstructed, their manual classification
as shower is questionable. Only 6 of the rejected events could be sucessfully reconstructed,
always yielding an energy below $\rm 10^{18}\,eV$, which is close to lower energy limit of the detector. 
 
Because the background rejection of our multiplicity trigger is already good enough at high 
pixel numbers, the space time correlation cut is only applied for events with less than 25
pixels in order to increase processing speed.

\section{Trigger performance}
After the cuts and algorithm of the new trigger had been optimized, we 
performed several tests in order to benchmark our
trigger and compare its performance with the prototype setup.

\subsection{Background rejection}
\label{sec:BG-rejection}
To verify the high background rejection efficiency, we used 
24306 events from randomly selected background files as an
independent data set. 
These background files are recorded during normal data taking
for monitoring purposes. 1\,\% of the events rejected by the
prototype TLT, and 10\,\% of the events rejected by the 
last trigger stage (T3) are stored. To get a realistic
composition of background events each prototype TLT rejected
event has to be weighted statistically with the factor 10.

Taking into account this weight factor we determine a background 
rejection of 95\,\% for the new TLT implementation (to be compared to
74\,\% for the prototype TLT).

As expected, the efficiency rises with the number of pixels,
99\,\% of the background events with more than 25 pixels 
(e.g. lightning) are rejected.

\subsection{Acceptance of simulated showers}
The extreme rareness of ultra high energetic cosmic
rays, limits their statistics in measured data.
In order to check the trigger acceptance to showers in the 
full primary energy range we used simulated events. The 
simulation was done with Offline \cite{offlinepaper} using CONEX-showers 
\cite{conexpaper} with energies between $10^{17}\,\rm eV$ and 
$10^{21}\,\rm eV$. Proton- and iron-induced showers were generated 
in equal parts, their positions were randomly distributed over the 
measuring field. The simulation took into account the fluorescence 
yield, light propagation and absorption, the properties of the telescope 
hardware, readout, FLT and SLT.

The fraction of falsely rejected showers (called \textbf{FRS} in the following) 
for the new TLT and the prototype are compared in table~\ref{tab:comparision_sim}. 
As some simulated showers are close to immanent detection limits due to large simulated 
distance and low primary energy, we also checked the events for reconstructability.
Events with sufficient data quality for reconstruction are accepted much better
by our TLT.

\begin{table}[ht]
    \centering
    \begin{tabular}{|l||c|c|}
      \hline
       & All events &  Reconstructable events \\\hline
	Number of events & 9257  & 5630 \\\hline 
	FRS, prototype TLT & 9.8\,\% &  6.1\,\%\\\hline 
	FRS, new TLT & 0.68\,\% &  0.18\,\% \\\hline
    \end{tabular}
\caption{Comparison of FRS with simulated showers for the prototype and new TLT.}
\label{tab:comparision_sim}
\end{table}

\subsubsection{Energy dependence}
Since the measurement of the energy spectrum is one of the main goals of the 
Pierre Auger Observatory, determination of the energy dependent 
trigger acceptance is essential. 

\begin{figure}[ht]
\centering
\includegraphics[width = 0.43\textwidth]{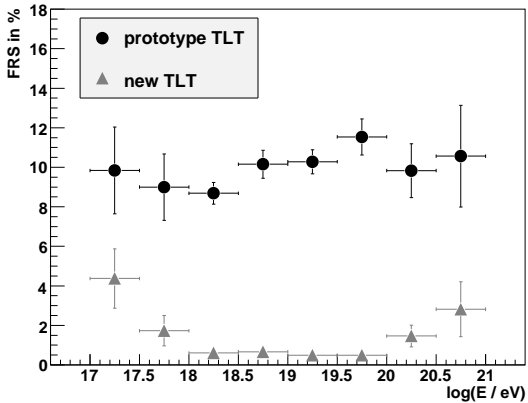}
\caption{Rejection of simulated showers
depending on their energy for the new and
the prototype TLT.}
\label{fig:acceptance_E}
\end{figure}

Figure~\ref{fig:acceptance_E} shows that the fraction of rejected
showers rises with energies below $\rm 10^{18}\,eV$ and above 
$\rm 10^{20}\,eV$. The higher FRS at low energies is simply explained by 
difficulties measuring dim showers close to the detector's energy threshold. 

Showers of very high energies 
are rejected for other reasons: The traces of near showers are
often so extended that they split up in two telescopes (multi-mirror event). 
The telescope viewing the small outermost part of the shower trace 
might reject this mirror-event with few pixels. In the highest energy-bin  
with 140 events, 2 of 3 rejected showers are such multi-mirror event 
tails. Of course the showers are not 
lost completely, as the neighboring telescope records the main part of the 
shower trace. To record the complete shower trace anyhow, an additional
inter-camera-trigger is under development. 

Allowing for this energy dependence and a particle flux 
$\Phi \propto E^{-3}$ the false shower rejection ratio in reality is 
dominated by the low energies to a value of 3.8\,\%. This result
is improved to 0.6\,\% when only taking into account showers with
$E > 10^{18}\,\rm eV$.

\subsubsection{Distance dependency}
As shower brightness and pulse 
width both depend on shower distance, this is one of the most important 
variables to parameterize for the detection probability. 

\begin{figure}[ht]
\centering
\includegraphics[width = 0.43\textwidth]{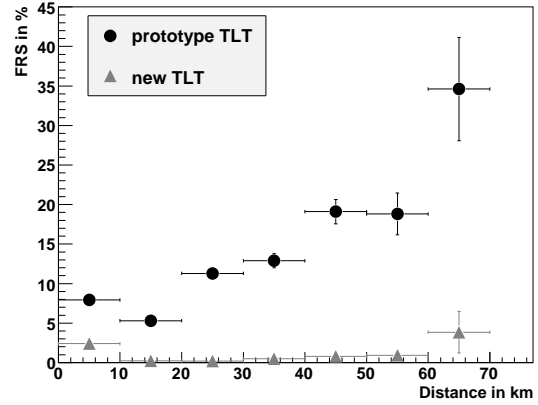}
\caption{Rejection of simulated showers
plotted versus their distance for the new and
the prototype TLT.}
\label{fig:acceptance_d}
\end{figure}

The rise of the FRS with shower distance is small up to 60 km for 
the new TLT, especially when compared with the prototype TLT (s.~fig.~\ref{fig:acceptance_d}).
Very near showers are also rejected more frequently, as they 
develop very fast and it becomes difficult to verify a temporal sequence 
within the limits given by the time resolution of the apparatus.

Other shower parameters like the kind of primary particle or the shower's 
direction were also examined and no trigger dependency was found.

\subsection{On site commissioning}
The real-time performance of the new trigger was tested during 5 nights 
of the August shift 2007 in one FD building. For 
direct comparison we operated telescope 1, 3 and 5 with the new and 
telescope 2, 4 and 6 with the prototype TLT. Thus weather and background
light conditions were comparable for both telescope sets.

The prototype and the new TLT achieved background rejection efficiencies of
66\,\% and 94\,\% respectively. These values are in good agreement with the 
offline estimations in section~\ref{sec:BG-rejection}.

Another goal of the new algorithm was to speed up the TLT decision.
Table~\ref{tab:cpu-consumption} summarizes the processing speed as measured
during commissioning. Especially background events with lots of pixels as they
occur in bursts during lightning are rejected several orders of magnitude faster.

\begin{table}[ht]
    \centering
    \begin{tabular}{|l||c|c|}
      \hline
        Decision time & prototype TLT &  new TLT \\\hline
       events,  $\rm n_{pix} \le 25$ & 25\,ms &  3.7\,ms \\\hline
       events, $\rm n_{pix} > 25$  & 109\,ms & 0.05\,ms\\\hline
       all events & 26\,ms & 3.5\,ms \\\hline
    \end{tabular}
\caption{Comparison of the average consumed CPU-time per event of both TLTs.}
\label{tab:cpu-consumption}
\end{table}

\section{Conclusions}
We have developed a new TLT software on the basis of the existing
fluorescence telescope hardware. The main goals of the development
were an improved background rejection and fast trigger decision 
in order to handle bursts of background events
during bad weather conditions.

These aims are achieved by using a two stage algorithm:
In a first step we use the temporal developement of the number of triggered
pixels in an event. This multiplicity signal is calculated in
hardware. Five independent cuts on multiplicity parameters are used to
reject background events. 

Although the multiplicity trigger works fine for the main data load of
events with many illuminated pixels, it fails in distinguishing short
shower traces from background events with less than 25 pixels. 
For these events we need a more detailed analysis of the temporal
appearance of the triggered pixels along the shower track.

The acceptance of extensive air showers was determined to be better than 99\,\%
using simulated events in the energy range from $10^{18}\,\rm eV$  to 
$10^{21}\,\rm eV$. At the same time 
95\,\% of the background is rejected, during bad weather conditions
the rejection efficiency of events with more than 25 pixels rises above
99\,\%. As the algorithm abstains from single pixel analysis for these events,
the decision is done fast enough to reject background in almost real-time. 

The new TLT was installed into the trigger chain in September 2007.
This enhancement improves the data quality of the fluorescence 
detector and noticeably simplifies data taking especially during the 
rainy season.

%Finally it should be remarked that the former lower trigger efficiency
%has no effects on the overall normalisation of the energy spectra of the
%Pierre Auger Observatory, because normalisation is taken only from the 
%surface detector.

\end{document}